# Aqueous solution interactions with sex hormone-binding globulin and estradiol: A theoretical investigation

A. J. da Silva[1,2] ● E. S. dos Santos[3]

**Abstract** Sex hormone-binding globulin (SHBG) is a binding protein that regulates availability of steroids hormones in the plasma. Although best known as steroid carrier, studies have associated SHBG in modulating behavioral aspects related to sexual receptivity. Among steroids, estradiol (17β-estradiol, oestradiol or E2) is well recognized as the most active endogenous female hormone, exerting important roles in reproductive and nonreproductive functions. Thus, in this study we aimed to employ molecular dynamics (MD) and docking techniques for quantifying the interaction energy between a complex aqueous solution, composed by different salts, SHBG and E2. Due to glucose concentration resembles those observed in diabetic levels, special emphasis was devoted to uncover the main consequences of this carbohydrate on the SHBG and E2 molecules. We also examined possible energetic changes due to solution on the binding energy of SHBG-E2 complex. In this framework, our calculations uncovered a remarkable interaction energy between glucose and SHBG surface. Surprisingly, we also observed solute components movement toward SHBG yielding clusters surrounding the protein. This finding, corroborated by the higher energy and shorter distance found between glucose and SHBG, suggests a scenario in favor of a detainment state. In addition, in spite of protein superficial area increment it does not exerted modification on binding site area nor over binding energy SHBG-E2 complex. Finally, our calculations also highlighted an interaction between E2 and glucose when the hormone was immersed in the solution. In summary, our findings contribute for a better comprehension of both SHBG and E2 interplay with aqueous solution components.

1,2 adjesbr@ufsb.edu.br or adjesbr@gmail.com
3 eliasss@ufba.br or silvasantosster@gmail.com

[1]Grupo de Pesquisa em Carcinologia e Biodiversidade Aquática (GPCBio)
[2]Centro de Formação em Ciências e Tecnologias Agroflorestais Universidade Federal do Sul da Bahia, Itabuna, Bahia, 45613-204, Brazil.

[3]Instituto de Física, Universidade Federal da Bahia, Campus Universitário de Ondina, Salvador, Bahia, 40210-340, Brazil.

# 1 Introduction

The representative effects of electrolytes on biomolecules have been rigorously explored since the Hofmeister pioneering studies on protein precipitation induced by salts [1]. In living organisms, huge family of pumps, channels and receptors regulates ionic levels [2]. Although the correspondent mechanisms are not fully understood, ions are known for changing solubility and melting temperature of proteins [3]. Specifically, relative to the endocrine system, several experimental investigations have revealed important biophysical aspects of the ionic interactions with hormones. For instance, Peggion *et al.* explained how calcium interact with gastrin family peptides in two different media made by trifluoroethanol solution and water, respectively [4]. It is also interesting to mention that Saboury *et al.* carried out a thermodynamic study on the interaction between magnesium and human growth hormone, showing a protein thermal stability increment [5]. Moreover, studies assessing modifications in electrolyte composition of cervical mucus provided a correlation between estrogen levels with chlorine and sodium [6]. On the other hand, despite the wealth of empirical studies involving ionic interactions with proteins, valuable information has also been harvested using simulations [7]. Theoretical methodologies as molecular dynamics (MD) and docking techniques have been vastly employed in recent years, thanks to the extensive development of analytical and numerical methods. These approaches, created for translating physical laws that represent biological systems, supply a faithful mechanistic representation of experimental results. Another remarkable advantage using *in silico* approach is the opportunity for preparing simulations based on difficult experimental maneuvers. For this reason, simulations emerge as a useful approach to provide insight into the intermolecular complex structure constituted by target and ligand, predicting both mode and free energy bindings by assuming scoring functions [8-9]. In this context, computer experiments were successfully used to investigate protein interactions with aqueous solution elements at different physicochemical conditions [10,11]. These simulations allowed better comprehension about sodium and chloride propagation near the lysosome surface, also elucidating the calcium role on dynamics and global structure of parvalbumin in aqueous solution [12,13]. Furthermore, computational investigations exposed zinc influences on SHBG-steroids binding properties, such as lysozyme properties modified by saline solution [14,15]. Besides their well-known interactions with ions, proteins forms covalent bonds with a variety of sugars in reactions known by glycation and glycosylation [16]. Both processes, commonly observed in metabolic derangements in diabetes, are characterized by high glucose concentration in the blood, differentiating from each other by absence (glycation) or presence (glycosylation) of enzymatic controlling. For example, an important effect of hyperglycaemic environment is the promotion of structural and functional modifications in serum albumin and myoglobin [17,18].

Daughaday firstly reported the existence of SHBG in the 50´s, but only after the investigations due to Mercier *et al.* the seminal findings were fully accepted [19]. SHBG binds to steroids in the plasma modulating the hormone availability in most groups of vertebrates including mammals, reptiles, amphibians and fishes [20]. Human SHBG is a homodimeric glycoprotein, produced and secreted by the liver, placenta and testes, where each monomer consists of 373 amino acids residues, existing as a 90 KDa protein containing a single binding site [21]. Crystallographic analysis at 1.55Å showed important details of steroid-binding site and quaternary structure of the dimer [22]. Recent reports have associated SHBG with other important physiological rules associated to sexual receptivity and metabolic disorders such as diabetes and obesity [21,23-26]. Among steroids bound to SHBG, we may highlight E2, which bind at dissociation constant in nanomolar range [27-28]. The molecule of E2 has low hydrophilic degree, molecular volume of 245$Å^3$ and apolar surface area of 261$Å^2$, being a final product of several enzymatic reactions including cholesterol and testosterone catabolism [29]. Since its discovery, many efforts have been devoted to elucidate the E2 physiological function, especially those related to molecular machinery of reproduction, enabling development of a wide range of drugs [27]. However, further investigation showed that E2 is not restricted to the peripheral reproductive system. In fact, E2 is related to diabetes, inflammation, and breast tumor, among others [30-33]. The brain is also a target for hormonal action synthesized from the neuronal metabolism. In this framework, E2 exerts notable influences in mechanisms relative to memory formation and neuroprotection, such as it modulates the dimorphism in brain morphology [34-36].

In resume, in this work our main purpose is to combine MD and molecular docking to investigate three main issues: (1) how dynamics and structural changes of SHBG are affected by electrolyte solution; (2) The glucose content adopted in the present work is within the diabetic range. This motivated us to quantify a possible role of glucose given by a possible cluster formation close to the protein; (3) potential solute interference on the SHBG-E2 binding energy.

# 2 Materials and Methods

## 2.1 Preparation of protein and ligand for docking

Calculations relative to the E2 docking on SHBG were based on three-dimensional structure of SHBG in complex with dihydrotestosterone (DHT) obtained from the Protein Date Bank (PDB code 1KDM). Hydrogen atoms were added and bonded using the Autodock Tools 4.2 with partial charges calculated using empirical force field MMFF94 [37]. In addition, binding site selected for docking was formed by the residues Thr40, Ser42, Phe56, Asp65, Phe67, Leu71, Leu80, Asn82, Val105, Met107, Val112, Ser128, Met139, and Ile141, Wat 364 and Wat 369. To supply new binding geometries, for both ligand and side chains, it was assumed all selected residues to be flexible. The selection was made manually based on knowledge of amino acids composing the active site. We obtained the ligand molecule (E2) from Pubchem database (pubchem.ncbi.nlm.nih.gov/) at the “sdf” file format. Open Babel Software version 2.3.1 was used to convert ligand representation into pdb format [38]. Next, we added all hydrogens and partial charges obtained applying the empirical force field MMFF94 [38]. Finally, we performed visual inspection through Pymol viewer version 1.7.x for correcting eventual structural errors on SHBG structure [39].

## 2.2 Simulation redocking

Simulations were performed using AutoDock Vina to the docking parameters adjustments [40]. We removed all water molecules from SHBG except those belonging the active site because they may occupy conserved positions by influencing in the ligand recognition process. We further extracted DHT substrate from SHBG, recoupling it again to the protein through redocking simulation. A simulation box was defined around the binding site of the protein whose center is under coordinates of DHT (-5.491, 39.238, 32.056), where box size is related to the directions x, y, and z are 50 Å. Better results in term of redocking were obtained comparing them with DHT crystallography position, as deduced by superposition of the two structures (Figure 1). Table 1 shows the best parameters found in redocking (RMSD 0.49 Å) applied to set E2 docking into SHBG.

## 2.3 Docking protein-ligand

We achieved E2 docking into SHBG using AutoDock Vina with all complexes visualized in PyMOL [39,40]. Briefly, the best binding modes procedure obeyed two main steps: (a) splitting of identical solutions in term of conformations, where results of low energy were separated for further visual comparison with DHT crystallographic structure position; (b) using both lower RMSD and energy values to generate the model to SHBG-E2 complex and its subsequent optimization with Steepest Descent method, allowing improvement in the accuracy of binding mode (Figure 2).

## 2.4 Molecular Simulation

Simulation was built using Gromacs 5.0.2 package adopting the Charmm27 force field with structure taken from the SHBG-E2 complex (Fig.2) obtained in the docking procedure [41]. Topology of E2 and salts were obtained from the server Web SwissParam with partial charges assigned of the force field MMFF94 [42]. To perform simulation, we manually extracted hydrogen atoms and water molecules from the complex. We next applied LINCS algorithm for all bonds within the complex allowing 2 fs time steps whereas SETTLE algorithm was adopted for water molecules bonds [43,44]. The cut-off to the non-bonded interactions was 2 nm and electrostatic interactions were computed using the Particle Mesh Ewald method [45]. We maintained temperature constant through the weak couple of V-rescale with time constant of 0.1 ps, being the pressure equilibrated at 1 bar through Parrinello-Rahman method employing a pressure of 2 ps time constant relaxation.

Our system considered SHBG and E2 immersed, at mammalian physiological temperature, into a saline solution where salts used are those found in Ringer physiological solution. Complex was soaked in an octahedron box of volume 712.03 $nm^3$ with water (700mM) TIP3P type, KCl (12.4 mM), $CaCl_2$ (27.1 mM), $MgCl_2$ (16.6 mM), $NaCO_3$ (44.5 mM), $Na_2H_2PO_4$ (16.8 mM), and glucose (24.5 mM) with margin of 1.2 nm between SHBG-E2 complex and each side of the box. We neutralized the system by adding 128 mM of NaCl molecules.

We have calculated the binding free energy between SBHG and E2 by LIE method by assuming two approaches. Calculations were done in SBHG-E2 complex followed by another considering E2 in a free protein aqueous media. Both systems (bound and free) were equilibrated in two phases: 5 ns in the NVT ensemble followed by the same time in the NPT ensemble. Free and bound systems were simulated during 300 ns, at temperature of 310 K and pressure of 1bar with initial velocity generated from the Maxwell–Boltzmann distribution. We calculated bound and free energies of molecular complex between a protein P and a ligand L by using:

$$\Delta G = G(P:L) - G(P) - G(L), \tag{1}$$

where ΔG represents the binding free energy difference between E2 bonded and non-bonded states. Simulation was performed on free E2 in the same way as described above for the SHBG-E2 complex to obtain the van der Waals and electrostatic interaction energies between E2 and surrounding environment. Empirical parameters used to calculate the free energy are α=0.348, and β=0.176 with average energies over the frames taken from all trajectories. Calculation of binding free energy was accomplished by using LIE method due to its faster performance in relation to other methods [46,47]. Briefly speaking, this method calculates van der Waals and electrostatic interactions relative to free and bound ligand, according to the following equation:

$$\Delta G_{bond} = \alpha\left(\left\langle V_{l-s}^{vdw}\right\rangle_{bond} - \left\langle V_{l-s}^{vdw}\right\rangle_{free}\right) + \beta\left(\left\langle V_{l-s}^{ele}\right\rangle_{bond} - \left\langle V_{l-s}^{ele}\right\rangle_{free}\right), \tag{2}$$

where the α and β parameters are van der Waals and electrostatic scaling factors, 〈 〉 denotes averages of the non-bonded van der Waals (vdw) and electrostatic (el) interactions between the ligand and surrounding environment (l-s), i.e. for both receptor binding site (bound state) and solvent (free state), respectively [48]. Finally, all protein figures were created using Pymol version 1.7.x and graphs Xmgrace computer program.

# 3 Results and discussion

## 3.1 Complex time evolution

Interaction between side chains and solvent is one of the factors responsible for the native structure of a protein. Consequently, solvent composition modification naturally changes the structural thermodynamics balance [49]. In this sense, due to the increment of pH or ionic concentrations, water molecules interact with ions by reducing the water-protein synergism. Thereafter, a decrement in protein-protein interactions induces structural alterations in the side chains of the binding sites and ligand affinity [50]. We analysed the root-mean-square deviation (RMSD) of the SHBG-E2 complex in relation to the optimized structure by inspecting the conformational states. Figure 3 shows conformational states evolution of complex from 0.11nm to 0.22 nm, where the first equilibrium state consolidates at 11.5-50 ns. After that, other partial states are observed from plateaus around 68-125 ns (0.20 nm), 200-227 ns (0.20 nm) and, 230-284 ns (0.22 nm), demonstrating several state transitions towards the final native state. This final configuration may still has some freedom of movement in the space dimensional, but may be sufficiently stable for its biological function [51]. However, it is necessary to take into account that physical models represent limited protein folding descriptions. In relation to E2, fluctuations around 0.47 nm indicates large displacements relative to the initial position. It is known that water molecules may contribute to ligand stabilization, allowing the ligand achieves active site due to side chains residues fluctuations. We have determined the presence of water molecules such as ions surrounding E2, using the radial distribution function:

$$g(r) = \frac{n(r)V}{N\Delta V}, \tag{3}$$

where $n(r)$ is the average number of atoms in a certain volume. Figure 4 illustrates a representative $g(r)$ of the ionic and water molecules distributions placed at distance $r$ from E2. Peaks in $g(r)$ indicates a larger solvent molecules density in the spherical shell at $r$=1.8nm, in relation to the ligand, corresponding to a well-structured solvation layer. Beyond this radius, there is a function vanishing, implying a null density due to repulsive van der Waals contribution. Furthermore, we did not find hydrogen bonds between water molecules or residues of the side chains and E2 (data not shown). Indeed, we have identified about three hydrogen bonds formation between H1 (E2) atoms with the CA atom (TRP66) and O1 (E2) with HC1 atoms (TRP60) and CA (ASP59), which contribute to E2 stabilization. For calculations related to hydrogen bonds we assumed $r\leq 0.35$nm and $\alpha = 30^{o}$, where $r$ is the minimum cut-off radius and $\alpha$ is the minimum angle required for hydrogen bonds formation. Absence of water molecules, salts or ions near to ligand assures us to admit that at least non-polar and electric interactions must be responsible to the SHBG-E2 binding.

## 3.2 Binding site

We suggest that conformational changes on SHBG are, at least partially, due to the salts over its surface. Undoubtedly, based on Figure 5, our simulation showed most ions exhibiting strong interaction with water molecules in the shallows SHBG. Additionally, it is also important to note a clear glucose cluster formed near the protein. Therefore, one could hypothesize that dynamic binding site and its interaction with E2 would produce changes in conformational states within the protein. Ionic compaction effect over SHBG was inspected on the accessible surface area to solvent (SASA), which increased from $80nm^2$ to $88nm^2$ during protein stabilization period (Figure 6). This SASA increment is a consequence of clustering formation on SHBG. Nonetheless, the same calculations relative to the SHBG binding site residues and E2 did not show any significant SASA modification during the simulation, maintaining itself in $26.3nm^2$ and $5.1nm^2$, respectively.

## 3.3 Energetic Analysis

In this section, we display the calculation of the energy associated to SHBG and E2 conformational changes given by the electrostatic (EL) and Lennard Jones (LJ) of potential energy. A resume of these calculations are presented in Table 2, where the electrostatic potential dominates interactions giving stabilization. However, LJ does occur in an opposite manner, at least for monovalent ions, but it is important to emphasize that in spite of some regions of surface captures ion and glucose it did not necessarily implied in salts motion decrement. This behaviour is credited to small modifications in atomic positions, strong enough for producing prominent fluctuations in LJ component that affects the potential stabilization. In this scenario, ionic forces on proteins evokes solubility decrement due to ionic competition with proteins for water molecules. Certainly, the present calculations highlighted a stronger SHBG interaction with glucose. This specific result motivated us to analyze the formation of glucose clusters on the protein in the last 60 ns of simulation. We addressed this issues based on Jarvis-Patrick algorithm, where molecules placed into a range of 0.2 nm, meaning that only the glucose molecules within this region are defined as belonging to the same cluster [52]. The cluster position was found to be apparently linked to its size, where the largest fraction of aggregate molecules is formed (on average) by 15 glucose molecules. We next quantified the interaction between brings results for simulations between E2 with aqueous solution components (Table 3). According these calculations it may be seen that binding of small ions to E2 is less intense when compared with glucose values. In addition, for this hormone, nonpolar contribution is the most important factor in the interaction with SBHG. Indeed, electrostatic interaction energy is estimated to be repulsive (0.102 kJ/mol) in relation to closest ions from E2. This behaviour reveals an interactive balance emerging as an important function in the folding states of binding site, where results indicate invariability of conformations, although it has been detected some points of great fluctuations on trajectories. Figure 7 brings the position of glucose (light blue) about E2, others ions such as $K^+$ (gray) and $CO_3$ (blue) are visualized and indicated by white arrows as well. Hydrophobic residues Phe32, Phe67 and, Tyr57 repel polar groups to effectively participate of the interactions with the E2 aromatic ring. However, to study the influence of hydrophobic term on the potential energy one must consider other aspects such as solvent cavitation energy among other factors. Table 4 show EL and LJ interaction potential components that composes ΔG calculated by using eq. (1). A non-polar term is associated to the ligand size in which apolar surface represents its hydrophobic capacity for effect with the binding site [53]. In this work, the ΔG obtained by LIE method was -54.45 ± 0.16 kJ/mol whereas the dissociation constant ($K_d$) obtained was $5.2 \cdot 10^{-10}$ M.

## 3.4 Further remarks

The present investigation aimed to characterize how aqueous solution interacts with SHBG and E2 as well as SHBG-E2 complex itself. To approximate of the physiological ambient we considered both physiological temperature such as a saline environment constituted by components found in the plasma of mammals. With this purpose, relative to the interplay of SHBG and aqueous constituents, our calculations provided for all ionic species distances ranged 0.5-1.2 nm from the protein surface (Table 3). Among solutes, glucose, $Na^+$ and $CO_3^{2-}$ showed the most pronounced energies and proximity to the SHBG surface. Solute assembly was already reported in solutions composed by sodium iodide dissolved in ether solution and aqueous NaCl solutions at high pressures and temperatures [54-57]. After both energy and distance stabilizing for glucose, $Na^+$ and $CO_3^{-2}$ these parameters remained unaltered during the rest of time scale simulation (data not shown). This result suggests the existence of a detainment state, already emphasized by previous investigations carried out on ion-residues system [58]. Furthermore, since our simulations were carried out during a certain period, this detainment might constitute an intermediary or temporary stage before glucose finally be able to form a covalent bond with SHBG molecule.

Plausible explanations for the glucose clusterization may be attributed to the formation of more hydrogen bonds (or saline bonds) and due to local density increment followed by preferential orientation of glucose molecules in the cluster in contrast to its ambient. Moreover, high energetic values for glucose, $CO_3^{2-}$ and $Na^+$ are explained by the partial water extinction from regions close to the SHBG with consequent hydrogen bonds formation and electrostatic interactions ruling solute interactions [59,60]. From a biological perspective, our results shows that glucose interaction with SHBG may emerges as additional mechanism to regulate freely available glucose in the plasma associating SHBG with type 2 diabetes [61]. Although their molecular mechanisms remains a conundrum, genetic evidence suggests that lower SHBG levels are observed in type 2 diabetes patients as compared to non-diabetic individuals [62]. Additionally, glucose assembly may also regulates SHBG conformational stability in response of other external perturbations [63-64]. In fact, Ohan and Dunn showed that glucose is a useful additive to stabilize collagen sterilized with irradiation [65]. In addition, Arakawa and Timasheff associated the area increment of different proteins, promoted by glucose, as the main responsible for the stability state [66]. This argument is congruent to the results obtained by Lins *et al.* considering interactions between trehalose with lysozyme protein in water solution at room temperature [67]. According to their results, trehalose moves toward protein by crowding around lysozyme surface due to water molecules partially removed away from the protein surface. Clustering was interpreted as an important element to promote lysozyme stabilization as well. In resume, glucose self-assembly may be a general mechanism for conformational stability enhancement of globular proteins against temperature changes. However, possible deleterious effects of high concentration of glucose and consequent crowding on SHBG are not discarded. This observation is influenced by previous work carried out with macromolecular aggregates composed by Dextran 70 and anionic protein lysate from *E.Coli* [68]. This issue could be assessed by considering glucose concentrations from the normal to diabetic range. Based on these observations, more investigation on the glucose concentration influence on SHBG is necessary to better characterize in more details the effects of these carbohydrate assembles.

Since E2 is partially soluble in aqueous solution, an important issue consists in identifying which solute component interacts with this hormone. In this framework, a theoretical inspection becomes especially relevant to understand the free but biologically active E2 in serum, where the hormone is exposed to the aqueous fluid while it moves toward a bind site. Our eenergetic analysis reveals that glucose is the main solute component interacting with E2 (0.102 kJ/mol and -0.322 kJ/mol for Coulomb and LJ, respectively). Based on this calculation is allowed to conclude that hormone diffusion within the aqueous solution was mainly influenced by attractive forces acting on E2-glucose interaction. We have also hypothesized if the aqueous medium interferes with the structural dynamics of SHBG, perturbing E2 binding with the substrate. Our results provided a $\Delta G$=-54.45 ± 0.16 kJ/mol for E2-SHBG complex, while other authors found $\Delta G$= -48.9 kJ/mol in experiments done in the human plasma [69]. Such differences in this result is attributed to the specific methodological procedure adopted in comparison with previous studies. Last, but not least, SASA calculations showed that clustering promotes increment of the surface SHBG area without affecting ligand or protein binding site areas. This result is theoretically supported by Ben-Naim hypothesis, which proposes that local modifications in protein folding, derived by the local peptides interactions, do not mean modifications in other areas of the protein [51]. From the experimental point of view, our calculations also converge to experimental investigations showing that carbohydrate interactions with SHBG did not affect steroids bind properties [70]. In addition, it is important to emphasize that the low affinity result obtained is particularly important to E2 transport because it facilitates the hormone releasing from the active site of SHBG to the estrogen receptor. Nevertheless, since our prediction is only based on simulations, experimental determination of ΔG is necessary to reinforce more accurately our theoretical values.

# 4 Conclusion

To the best of our knowledge, this is the first report to combine MD and molecular docking considering an ionic environment, which mimics the salts found physiological solution constituents, quantifying the energetic aspects of the interaction of electrolyte solution with SHBG and E2. We are convinced that the present work contributes for a better understanding of energetic aspects related to the E2-SHBG-salt interaction in a more complex fluidic environment. However, further work is necessary to determine specific SHBG residues site or attractor domains that interact with glucose and ions. Furthermore, other relevant issues consists to examine how glucose cluster is affected by both ionic concentration, pH and temperature changes. Lastly, in order to achieve a more realistic physiological scope, our future studies will consider salt concentrations completely compatible to the Ringer solution.

**Acknowledgments** Special thanks to prof. Márcio R. Maia and prof. Danielle O. C. Santos for valuable comments and suggestions during the preparation of the manuscript.

# Conflict of Interest

The authors declare no conflict of interest.

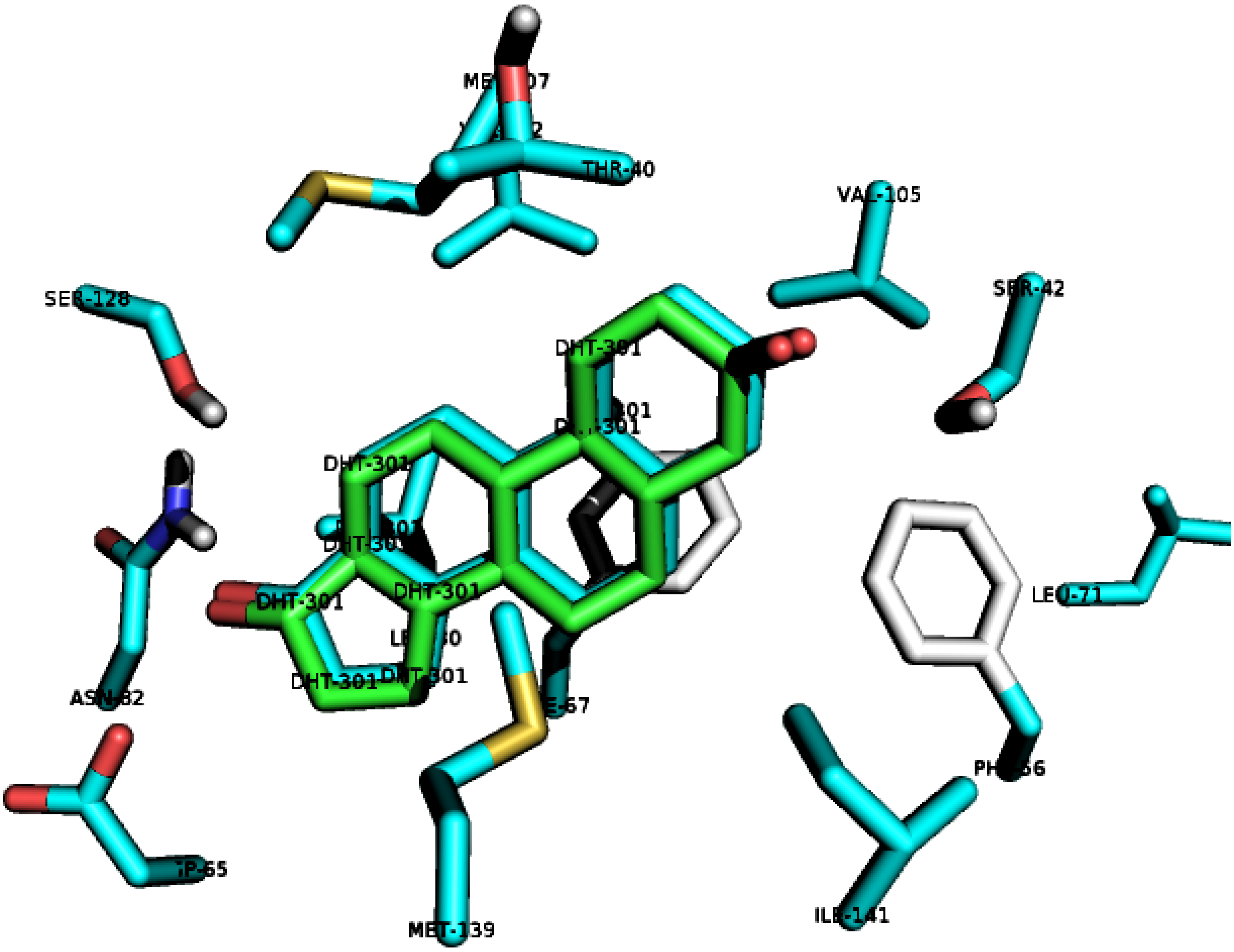

**Fig. 1** Overlapping of the DHT in redocking simulation. Docked DHT is coloured in green and red. Image created using Pymol Software

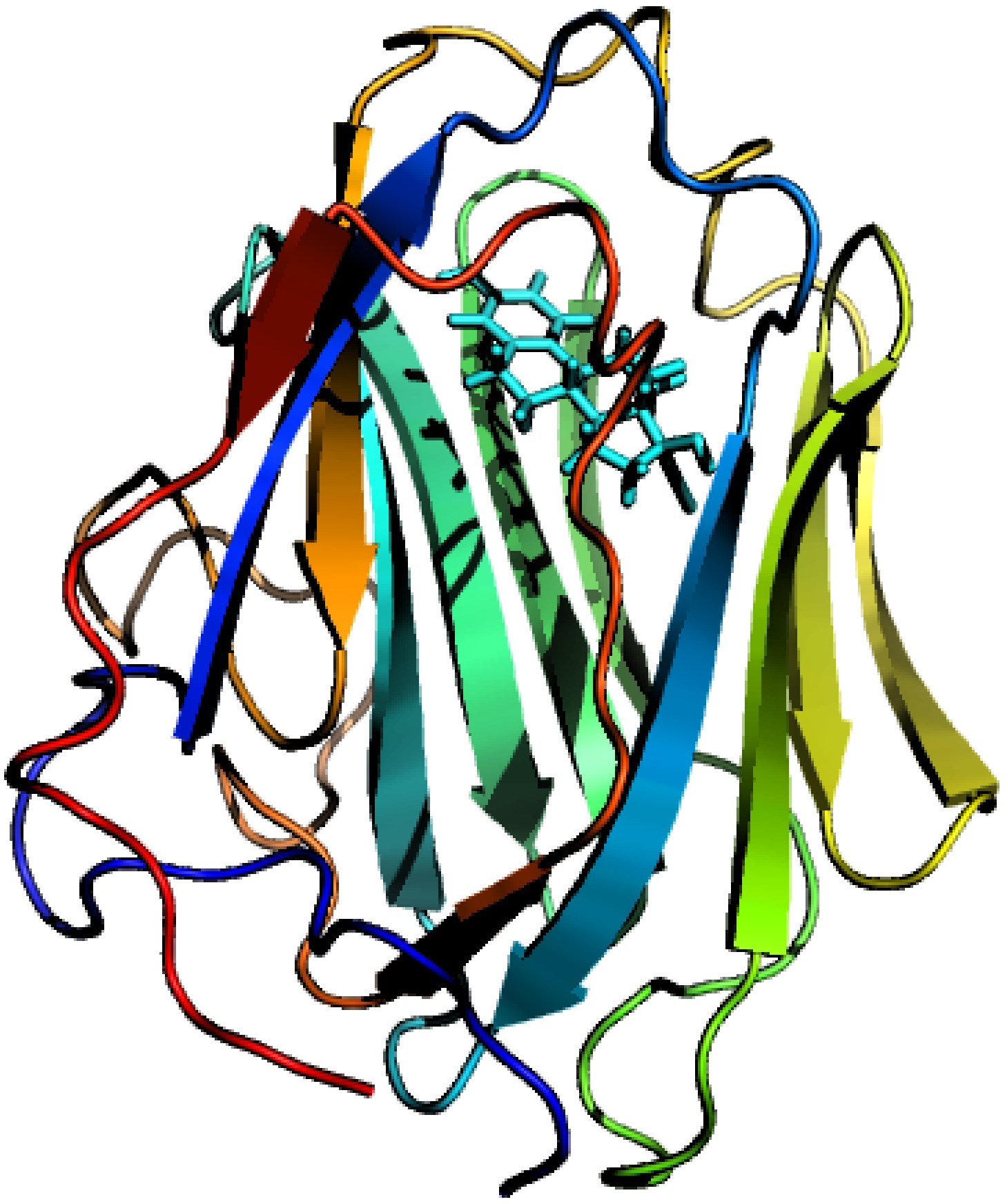

**Fig. 2** 3D-Estructure of SHBG. Image created using VMD Software

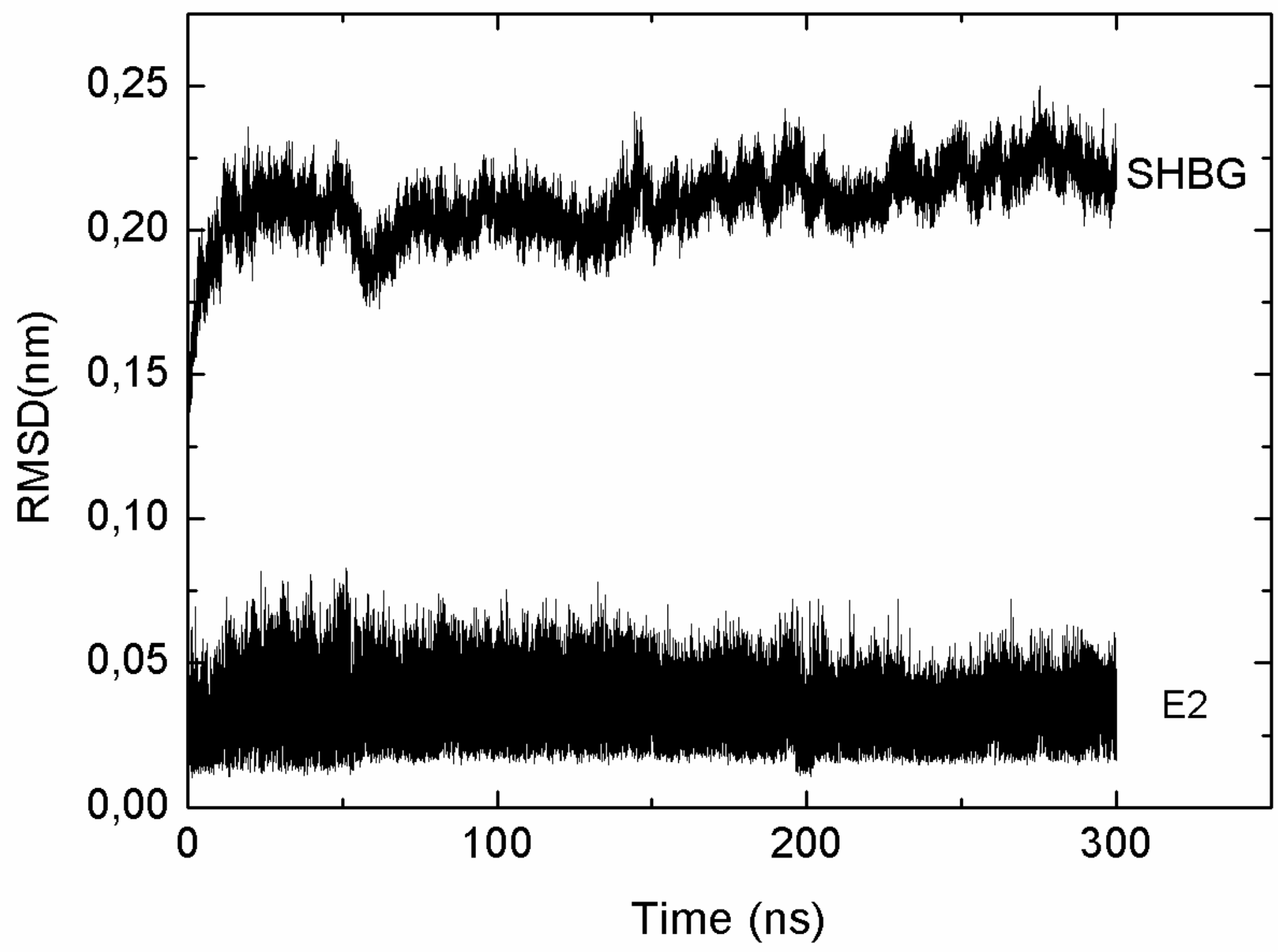

**Fig. 3** Root mean square deviation-RMSD of SHBG and E2

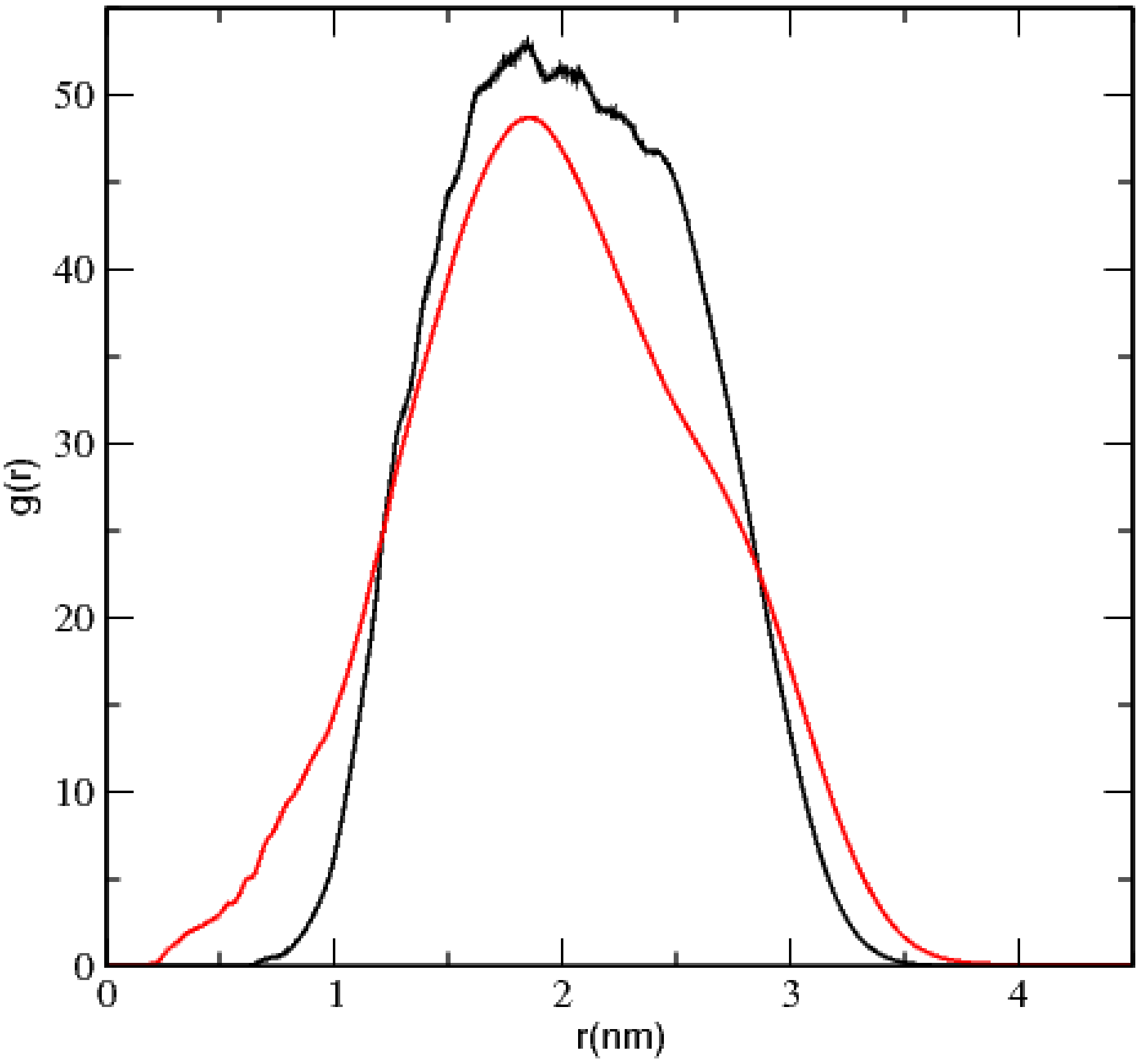

**Fig. 4** Radial distribution function of shells of ions (black) and water (red) surrounding E2

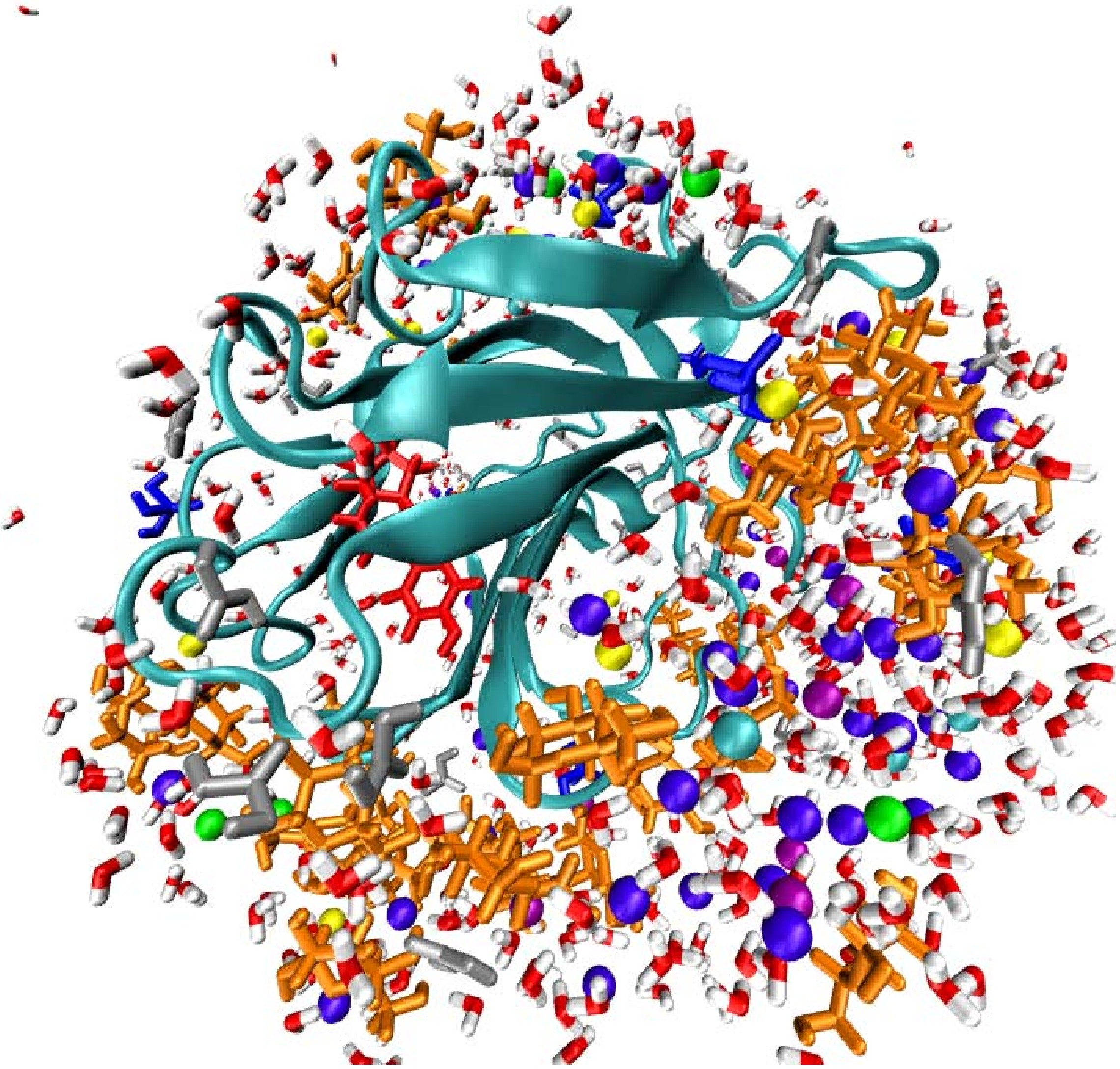

**Fig. 5** Representative snapshot of simulation at 310 K, where ions were represented as spheres and presents a magnified size for clarity. Water molecules are shown as bonds (red/white), $Na^{+}$ (yellow), $Ca^{2+}$ (purple), $K^{+}$ (green), $Mg^{2+}$ (cyan), $Cl^{-}$ (violet). $CO_3^{-2}$ (silver), $PO_4^{-2}$ (blue), and glucose (orange) as sticks. E2 molecule is represented as sticks colored in red. Image was created by using VMD Software

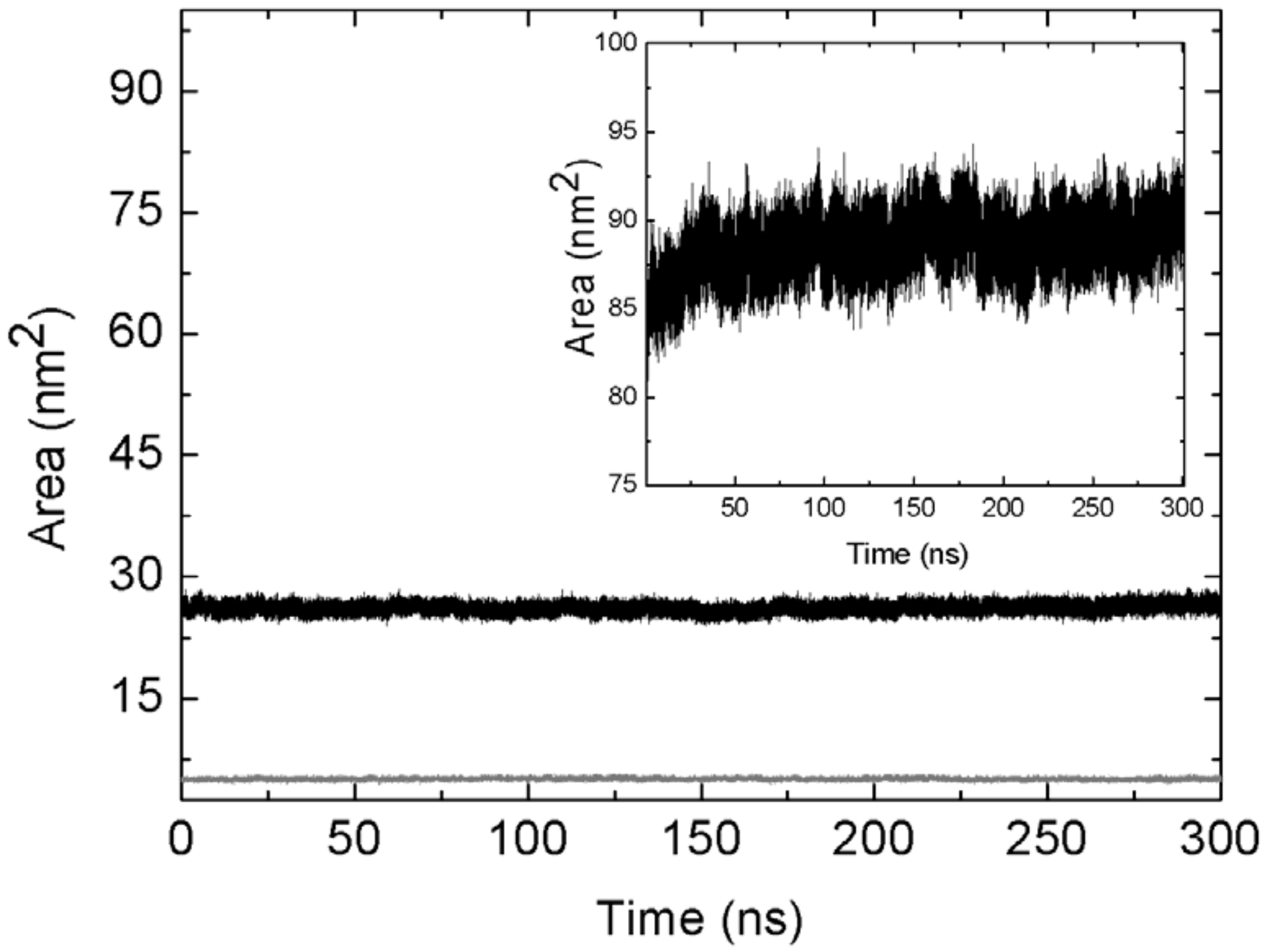

**Fig. 6** Fluctuation of solvent accessible area (SASA) for binding site (black) and E2 (grey). The inset shows SASA fluctuation of protein

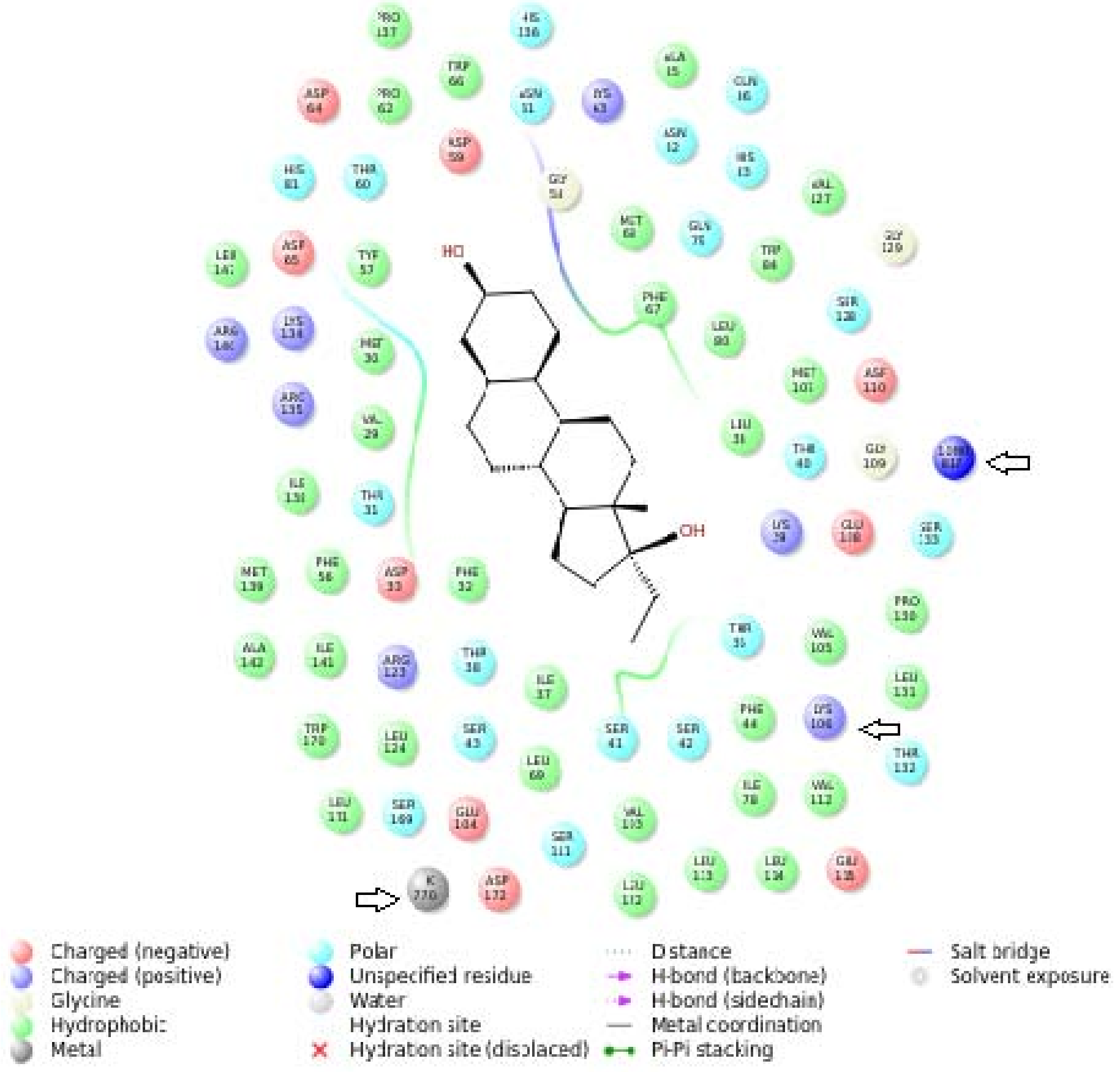

**Fig. 7** Interactions between E2 and SHBG created with Maestro Software. The arrows indicate the closest ions $K^+$ (grey), glucose (light blue) and, $CO_3^{-2}$ (blue)

| Parameters | Value |
|---|---|
| Energy range | 10 |
| Num modes | 20 |
| Exhaustiveness | 8 |
| Seed | -150 |

**Table 1** Parameters found in redocking used to set E2 docking into SHBG. Energy range describes the maximum energy difference between the best and worst binding mode; "Num modes" is the parameter associated to the maximum number of binding modes to be generated; "Exhaustiveness" is related with thoroughness of search, roughly proportional to time; "Seed" is associated to a random number generator

| Energy (kJ/mol) | $Ca^{2+}$ | $Cl^{-}$ | $K^{+}$ | $Mg^{2+}$ | $Na^{+}$ | $CO_3^{2-}$ | $PO_4^{-2}$ | Glucose |
|---|---|---|---|---|---|---|---|---|
| Coulomb (SR) | -1038.5 | -607.4 | -236.6 | -733.5 | -1524.3 | -1292.7 | -402.55 | -1474.5 |
| LJ (SR) | 106.60 | 4.82 | 20.45 | 47.42 | 154.68 | -158.92 | -123.35 | -624.45 |
| Distance (nm) | 1.2 | 0.9 | 1.0 | 1.3 | 1.0 | 0.7 | 1.0 | 0.5 |

**Table 2** Electrostatic and Lennard-Jones interactions between SHBG and aqueous solution components. Standard deviations were omitted due to their very small values

| Energy(kJ/mol) | $Ca^{2+}$ | $Cl^{-}$ | $K^{+}$ | $Mg^{2+}$ | $Na^{+}$ | $CO_3^{2-}$ | $PO_4^{-2}$ | Glucose |
|---|---|---|---|---|---|---|---|---|
| Coulomb (SR) | $2.67x10^{-8}$ | $2.3x10^{-4}$ | $7.4x10^{-4}$ | $2.83\ x10^{-8}$ | $-2.84x10^{-5}$ | $-1.92x10^{-2}$ | $-7.34\ x10^{-6}$ | $1.02x10^{-1}$ |
| LJ (SR) | $-5.0x10^{-8}$ | $-52.0x10^{-4}$ | $-35.65x10^{-4}$ | $-3.84\ x10^{-9}$ | $-26.01x10^{-5}$ | $-13.36x10^{-2}$ | $-3.81x10^{-5}$ | $-3.22x10^{-1}$ |
| Distance (nm) | 0.2 | 0.2 | 0.25 | 0.19 | 0.21 | 0.15 | 0.16 | 0.14 |

**Table 3** Electrostatic and Lennard-Jones interactions between E2 and aqueous solution components. Standard deviations were omitted due to their very small values

| Energy(kJ/mol) | $\langle V_{bond}^{vdw} \rangle$ | $\langle V_{free}^{vdw} \rangle$ | $\langle V_{el}^{bond} \rangle$ | $\langle V_{el}^{free} \rangle$ | $\Delta G$ |
|---|---|---|---|---|---|
| E2 | -146.68 ± 0.90 | -4.69± 0.39 | -69.99± 1.70 | -41.40± 1.80 | -54.45± 1.10 |

**Table 4** Average of the bind free energy components for E2. Results are expressed as mean ± standard deviation